\documentclass[12pt]{article}
\usepackage{amssymb,amsmath,epsfig}

\begin{document}

\title{\bf Cold Plasma Wave Analysis in Magneto-Rotational Fluids}

\author{M. Sharif $^1$\thanks{msharif@math.pu.edu.pk} and Umber Sheikh $^2$\\
$^1$ Department of Mathematics, University of the Punjab,\\
Quaid-e-Azam Campus Lahore-54590, Pakistan.\\
$^2$ Department of Applied Sciences,\\
National Textile University Faisalabad-37610, Pakistan}
\date{}
\maketitle

\begin{abstract}
This paper is devoted to investigate the cold plasma wave
properties. The analysis has been restricted to the neighborhood
of the pair production region of the Kerr magnetosphere. The
Fourier analyzed general relativistic magnetohydrodynamical
equations are dealt under special circumstances and dispersion
relations are obtained. We find the $x$-component of the complex
wave vector numerically. The corresponding components of the
propagation vector, attenuation vector, phase and group velocities
are shown in graphs. The direction and dispersion of waves are
investigated.
\end{abstract}
{\bf Keywords:} 3+1 formalism, GRMHD, Kerr planar analogue, cold
plasma, normal and anomalous dispersion.\\
\\
{\bf PACS numbers:} 95.30.Qd, 95.30.Sf, 04.30.Nk

\section{Introduction}

Einstein's general theory of relativity describes gravity as a
curvature of spacetime caused by the presence of matter. Most
massive physical objects like black holes distort the geometry of
spacetime by their immense gravity. The accretion disk around a
rotating black hole (most physical in nature) is always filled
with charged plasma due to the pair production in this region.
This charged plasma creates an external magnetic field around the
black hole (the region containing this field is termed as
magnetosphere in literature) and is governed by the general
relativistic magnetohydrodynamical (GRMHD) equations. The intrinsic
angular momentum of the black hole intends the surrounding plasma to
act as a magneto-rotational fluid. The black hole spin modifies
the rotation of the plasma which flows in the form of an
MHD wind. When this wind is perturbed analytically, GRMHD waves
are formed. These waves propagate to transmit information in the
magnetosphere. They also help us to detect whether the energy
extraction is allowed by the plasma state in the magnetosphere.

There is a large body of literature available \cite{RW}-\cite{MK}
about the perturbations and GRMHD waves in the regime of the
Schwarzschild black hole. The study of rotating black holes has
been an interesting subject since they lie in the hearts of active
galaxies \cite{L}. Blandford and Znajek \cite{BZ} discussed the
possibility of an electromagnetically driven wind from the
rotating black hole. Takahashi et al. \cite{T1} discussed the MHD
inflow and the extraction of energy in the form of jets.

In general relativity, different observers generally experience
distinct time measures. In 1962, an attempt was made by Arnowitt
et al. \cite{ADM} who quantized the gravitational field by a
specific foliation (now called ADM 3+1 formalism). Thorne and
Macdonald \cite{TM1}-\cite{TPM} used the ADM 3+1 split to bridge
the gap between the black hole electrodynamics and that of flat
space electrodynamics. The formalism was taken over by Holcomb and
Tajima \cite{HT} and Dettmann et al. \cite{De} to discuss the
quantum description of plasma including electromagnetic waves in
FRW metric. Buzzi et al. \cite{BHT1} used this access to
investigate the wave phenomenon near the Schwazschild event
horizon by using two component plasma. Recently, Sharif and Sheikh
\cite{S1}-\cite{S5} worked on the properties of cold and
isothermal plasmas living in the neighborhood of the Schwarzschild
event horizon. Komissarov \cite{Ko1} discussed Blandford-Znajek monopole solution
for black hole electrodynamics using 3+1 formalism. Khanna
\cite{Kh} derived the generalized Ohm's law in 3+1 split.

The energy can be extracted from the ergosphere via a process
first proposed by Penrose \cite{Pen}: a particle entering the
ergosphere can split in two in such a way that one fragment
falls into the hole, but the other leaves the ergosphere with
more energy than the original particle. The extra energy
comes from the hole itself. Leiter and Kafatos \cite{LK} discussed
the Penrose pair production in massive Kerr black hole's
ergosphere. Kafatos \cite{Kaf} discussed the gamma ray observations
from Penrose powered black holes. Koide et al. \cite{Ko} modeled the
GRMHD behavior of plasma flowing into the rotating black hole
in a magnetic field and showed (using numerical simulations)
that energy of the spinning black hole can be extracted
magnetically. Another way of extracting black hole's spin energy
is the form of super-radiance. Kokkotas and Schmidt \cite{KS} discussed
quasi normal modes for black holes and relativistic stars. They
have also discussed the w-waves or gravitational wave modes in
these scenarios. Furuhashi and Nambu \cite{FN} discussed the
super-radiance in charged Kerr spacetime by numerically solving the Klein-Gordon
equation as an eigen value problem. Zeldovich \cite{Zeld} considered the waves
scattering by a rotating conducting cylinder. They proved that Kerr geometry
displays super-radiance. Zhang \cite{Z1}-\cite{Z2} formulated
the black hole theory for stationary symmetric GRMHD with its
applications in Kerr geometry. He showed (taking the specific
values of the angular frequency and the $x$-component of the wave
vector) that the cold plasma allows the outflow of energy flux.
Recently, Sharif and  Sheikh \cite{S6}-\cite{S10} extended this
work to cold and isothermal plasmas by investigating the real wave
numbers. They verified the results given by Zhang \cite{Z2}. This
work partially helps to answer the question of energy extraction.

In this paper, we shall discuss the cold plasma wave properties by
calculating complex wave numbers. This provides an alternative to
verify the results obtained in \cite{S10}. We have considered the
Fourier analyzed perturbed GRMHD equations for the planar analogue
of the Kerr metric. These equations lead to the dispersion
relation and consequently the $x$-component of the complex wave
vector for the cold plasma. This complex $x$-component can be
split into $x$-components of the propagation factor (real part)
and attenuation factor (imaginary part). Further, we evaluate the
$x$-components of the phase and group velocities etc. near the
pair production region. These quantities would help us to
understand the modes of dispersion.

The layout of this paper is as follows. Section \textbf{2}
provides the description of spacetime near the pair production
region with some assumptions. Section \textbf{3} consists of the
background flow assumptions and related quantities used in
specification and interpretation of results. In Section
\textbf{4}, we Fourier analyze the GRMHD equations for the cold
plasma. Section \textbf{5} is devoted to the numerical solutions
of the dispersion relation. It also includes figures and related
discussion. The observable quantities are given which help us to
judge the waves properties more efficiently. We shall conclude
this discussion in Section \textbf{6}.

\section{Description of Model Spacetime}

The most general line element in 3+1 formalism can be written as
\begin{equation}\label{st1}
ds^2=-\alpha^2dt^2+\gamma_{ij}(dx^i+\beta^idt)(dx^j+\beta^jdt),
\end{equation}
where $\alpha$ is the lapse function, $\beta^i$ are the components
of shift vector and $\gamma_{ij}$ are the components of spatial
metric. All these quantities are functions of time and space
coordinates.

We consider the planar analogue of the Kerr metric \cite{S6}
\begin{equation}\label{st2}
ds^2=-dt^2+(dx+\beta(z)dt)^2+dy^2+dz^2.
\end{equation}
The directions $z$, $x$ and $y$ are the analogue of the Kerr's
radial $r$, axial $\phi$ and poloidal $\theta$ directions
respectively whereas $t$ represents the time coordinate. Utilizing
coordinate freedom, the lapse function is set to unity without
loss of generality. This has been done to avoid complications
which arise due to the presence of horizon and hence, the
red-shifts. The rotation of the black hole imposes a shift vector
which is obvious from Eq.(\ref{st2}). The value of the shift
function $\beta$ (analogue to the Kerr-type gravitomagnetic
potential) decreases monotonically from $0$ ($z \rightarrow
\infty$) to some constant value ($z \rightarrow-\infty$). Equation
(\ref{st2}) shows that our assumed $\beta$ is $\beta(z)\equiv
\beta(z)\textbf{e}_\textbf{x}$ which derives an MHD wind and
extract translational energy analogous to the rotational energy of
the Kerr metric. The plasma particles, created at $z=0$ (pair
production region), are then driven up to relativistic velocities
by magnetic-gravitomagnetic coupling, as they flow off to
"infinity" $(z=+\infty)$ and down towards the "event horizon"
$(z=-\infty)$.

Our scenario is based on the pair production region and its surrounding
plasma. Thus we are interested in the magnetospheric plasmas rotating in the
2-D fashion i.e. in ($zx$-plane). The shift is restricted to be dependent
on only one dimension. It may depend on the other dimensions which obviously
make the analysis tough but more general. The shift is the effect of black
hole rotation, of course. The plasma is being produced in the pair production
region (electron-positron pair production holds and at $z=0$ in our scenario)
outflows from the region towards the event horizon which lies at $z=-\infty$
and towards the outer end of the magnetosphere (the end away from the event
horizon). The plasma produced in the pair
production region falls out from the region. This outflow disturbs the neighboring magnetospheric plasma in
the form of waves.  The region $z=0$ is not a physical
singularity but the region where particle-antiparticle pairs are produced
from the vacuum energy. Since the out flowing plasma is moving
in the $xz$-plane, so it will absolutely disturb the outer plasma in the same
plane.

This spacetime requires a fiducial reference observer (FIDO)
(analogous to the zero angular momentum observer of the 3+1 split
of the Kerr spacetime), the one with four-velocity perpendicular
to the hypersurfaces of constant time $t$. The geometrized units
will be used throughout the paper.

\section{Plasma Flow and Relative Assumptions}

We consider a rotating background which is assumed to be filled
with cold plasma admitting the following equation \cite{Z2}
\begin{equation}\setcounter{equation}{1}
\label{cp} \mu=\frac{\rho}{\rho_0}=constant,
\end{equation}
where $\mu,~\rho_0$ and $\rho$ are the specific enthalpy, rest and
moving mass densities of the fluid respectively. This shows that
the plasma has vanishing thermal pressure and thermal energy. Thus
the cold plasma particles are moving along timelike geodesics of
the spacetime (\ref{st2}).

The rotating fluid's four-velocity and magnetic field measured by
our FIDO can be described by spatial vector fields lying in
$xz$-plane
\begin{eqnarray}
\textbf{V}=V(z)\textbf{e}_\textbf{x}+u(z)\textbf{e}_\textbf{z},\quad\
\textbf{B}=B\{\lambda(z)\textbf{e}_\textbf{x}
+\textbf{e}_\textbf{z}\},
\end{eqnarray}
where $B$ is a constant.

For the stationary symmetric background, the relationship between
$x$ and $z$ components of the velocity vector for the spacetime
(\ref{st2}) can be written as \cite{Z2}
\begin{equation}\label{xz}
V=C+\lambda u,
\end{equation}
where $C\equiv\beta+V_F$ is the normalized shift and $V_F$ is an
integration constant. Further, the mass conservation law in
three-dimensional hypersurface is
\begin{equation}\label{clm}
\alpha \rho_0\gamma u=A~(constant),
\end{equation}
where $\gamma=\frac{1}{\sqrt{1-u^2-V^2}}$ is the Lorentz factor.

We take the assumptions $V_F=1,~\beta=\tanh(z)-1$ for
simplifications. It is important to note that our assumed $\beta$ is
continuous across $z=0$ (i.e., the pair production region). This
shift may not be valid for the near horizon. The expansion of shift
vector $\beta$ implies that this vector is directly proportional to
the dimensionless angular momentum $J=\frac{a}{M}$ of the black hole
and inversely proportional to the radius of the black hole. Thus the
spin of the black hole can be expressed by the following relation
\cite{TPM}
$$\beta=\frac{2JM}{(r')^3}(ye_x-xe_y)+O(\frac{1}{r'^4}),$$ where
$r'=\sqrt{x^2+y^2+z^2}.$ This $\beta$ is involved in all the
dispersion relations. As the function $\beta$ changes, the value
of the dispersion relation (i.e., the relationship between the
wave number and angular frequency) changes. In this work, we
consider only the propagation which changes due to change in $z$
and hence the change in $\beta$.

According to these assumptions, the MHD flow is intended to model
a wind flowing out of the ergosphere of a Kerr black hole and also
an accretion flow down to the horizon, both originated at the pair
production region, i.e., $z=0.$ At the pair production region,
$\beta=-1.$ The renormalized shift vector is $C=\beta+V_F$ where $
V_F=1,$ is a flow constant. There is no flow at $z=0,$ thus
$C=\beta+V_F =-1+1=0$ at $z=0.$

Our assumed plasma's magnetic field having a variable $x$-component
$\lambda B$ with variable $\lambda$ depends on the variable $z$. The
plasma is frozen-in to the $B$-field and we have assumed an
adiabatic flow. For simplification, we have also taken $\lambda=1,$
i.e., the $x$-component of the magnetic field becomes constant and
leads to a constant unperturbed magnetic field. When we apply these
values to Eq.(\ref{xz}) and then substitute the resulting value of
$V$ in Eq.(\ref{clm}) with the simplification $A/\rho_0=1$ (as done
in \cite{S6}-\cite{S9}), we obtain the following pairs of velocity
components.
\begin{eqnarray}\label{u1}
&&u_1=-\frac{1}{3}\tanh(z)-\frac{1}{3}l,~V_1=\frac{2}{3}\tanh(z)-\frac{1}{3}l,\\
\label{u2}
&&u_2=-\frac{1}{3}\tanh(z)+\frac{1}{3}l,~V_2=\frac{2}{3}\tanh(z)+\frac{1}{3}l,
\end{eqnarray}
where $l=\sqrt{3-2\tanh^2(z)}$. The corresponding Poynting vector,
then, takes the form
\begin{equation}
\textbf{S}=\frac{1}{4\pi}\textbf{E}\times\textbf{B}
=2\tanh(z)(\textbf{e}_\textbf{x}-\textbf{e}_\textbf{z}).
\end{equation}
These assumptions lead to a rotating black hole magnetospheric
fluid which admits a constant external magnetic field. The fluid
is moving with a velocity $\textbf{V}$ in the magnetosphere. The
gravitomagnetic waves and pair of particles are produced in the
$z=0$ region. The energy extraction from the black hole is
possible if the medium allows the waves and particles to pass
through it and move out of the magnetosphere. The gravitomagnetic
waves transform information within plasma. If the medium, in the
region under consideration, allows the waves to move normally,
these waves can get out of the magnetosphere. This can be well
understood by investigating properties of the waves in this
region.

The plasma flow in the magnetosphere is perturbed due to gravity
and rotation of the black hole. This flow is characterized by
fluid's density $\rho$, velocity $\textbf{V}$ and magnetic field
$\textbf{B}$. The linear perturbations will be in $xz$-plane and
thus are dependent on $x,~z$ and $t$. The perturbed variables take
the following form
\begin{equation}\label{pv}
\rho=\rho^0+\rho\tilde{\rho},\quad
\textbf{V}=\textbf{V}^0+\textbf{v},\quad
\textbf{B}=\textbf{B}^0+B\textbf{b},
\end{equation}
where unperturbed quantities are denoted by the superscript zero.
Further, the dimensionless notations $\tilde{\rho},~\textbf{v}$
and $\textbf{b}$ can be expressed as
\begin{eqnarray}\label{p}
\tilde{\rho}&\equiv&\frac{\delta \rho}{\rho}=\tilde{\rho}(t,x,z),\nonumber\\
\textbf{v}&\equiv& \delta
\textbf{V}=v_x(t,x,z)\textbf{e}_\textbf{x}+v
_z(t,x,z)\textbf{e}_\textbf{z},\nonumber\\
\textbf{b}&\equiv& \frac{\delta
\textbf{B}}{B}=b_x(t,x,z)\textbf{e}_\textbf{x}
+b_z(t,x,z)\textbf{e}_\textbf{z}.
\end{eqnarray}

The perturbations are assumed to be harmonic, i.e., of the form
$\delta=e^{-i(\omega t-k_xx-k_zz)}$ (having sinusoidal dependence on
$t,~x$ and $z$).
\begin{eqnarray}\label{ps}
\tilde{\rho}(t,x,z)=c_1\delta,~
v_x(t,x,z)=c_2\delta,~v_z(t,x,z)=c_3\delta,\nonumber\\
b_x(t,x,z)=c_4\delta,~b_z(t,x,z)=c_5\delta,
\end{eqnarray}
where $c_1,~c_2,~c_3,~c_4,~c_5$ are arbitrary constants, $k_x$ and
$k_z$ are the $x$ and $z$-components of the wave vector
$\textbf{k}=(k_x,0,k_z)$ and $\omega$ is angular frequency of the
wave.

\section{Fourier Analyzed Perturbed GRMHD Equations}

The perfect GRMHD equations for the Kerr planar analogue
(Eq.(\ref{st2})) is given by Eqs.(2.4)-(2.8) of \cite{S7}.
Substituting the value of rest-mass density from Eq.(3.1), we
obtain their specific form for the cold plasma. The insertion of
perturbed values from Eq.(\ref{pv}) and then from Eq.(\ref{p})
give the component form of these equations. This is the same
procedure as given in \cite{S1} and \cite{S6}.

Since the perturbed variables have harmonic dependence on $x,~z$
and $t$, hence Eq.(\ref{ps}) is used and the following Fourier
analyzed form is obtained.
\begin{eqnarray}{\setcounter{equation}{1}}\label{a4}
&&\iota k_z c_2-(\iota k_z\lambda+\lambda')c_3-c_4(\iota k_z
u-\iota\omega+u')+c_5(m\iota k_z+m')=0,\\
\label{d4} &&\iota k_x c_2-\iota k_x\lambda c_3+\iota c_5
\{mk_x+k_zu-\omega\}=0,\\\label{c4}
&&k_xc_4=-k_zc_5,\end{eqnarray}
\begin{eqnarray} \label{e4}
&&c_1\iota\{-\omega+mk_x+uk_z\}+c_2\left[-\iota w\gamma^2V+\iota
k_z\gamma^2uV+\iota
k_xq+\right.\nonumber\\
&&\left.\gamma^2u\{(1+2\gamma^2V^2)V'+2\gamma^2uVu'\}\right]+c_3\left[-\iota(\omega+\beta
k)\gamma^2u\right.\nonumber\\
&&\left.+\iota k_zr+\iota k_x\gamma^2uV+2\gamma^4u^2VV'-(1-2\gamma^2u^2)r\frac{u'}{u}\right]=0,\\
\label{f4} &&c_1\rho\gamma^2u\{qV'+\gamma^2uVu'\}+c_2\left[-\iota
w\left\{\rho\gamma^2q+s\right\} +\iota
y\left\{\rho\gamma^2q-s\right\}\right.\nonumber\\
&&\left.
+\rho\gamma^4u\{(1+4\gamma^2V^2)uu'+4qVV'\}\right]+c_3\left[-\iota
w\left\{\rho\gamma^4uV-\lambda s\right\}\right.\nonumber\\
&&\left.+\iota y\left\{\rho\gamma^4uV +\lambda
s\right\}+\rho\gamma^2[\{(1+2\gamma^2u^2)(1+2\gamma^2V^2)
\right.\nonumber\\
&&\left.-\gamma^2V^2\}V'+2\gamma^2(1+2\gamma^2 u^2)uVu'] +u
\lambda's\right]+sc_4\{-\iota k_z(1-u^2)\nonumber\\
&&+uu'\}+sc_5\left\{-\lambda'-um'+\iota k_x(1-V^2)-2\iota
uVk_z\right\}=0,\\\label{b4}
&&c_1\gamma^2\rho[u\{ru'+\gamma^2VuV'\}-V\beta'] +c_2\left[-\iota
w\left\{\rho\gamma^4uV-\lambda s\right\}\right.\nonumber\\
&&\left.+\iota y\left\{\rho\gamma^4uV +\lambda s\right\}
+\rho\gamma^2\{\gamma^2u^2V'(1+4\gamma^2V^2)-\beta'(1+2\gamma^2V^2)\right.\nonumber\\
&&\left.+2V\gamma^2uu'(1+2\gamma^2u^2)\}\right]
+c_3\left[-\iota\left\{\rho\gamma^2(1+\gamma^2
u^2)+\lambda^2 s\right\}w\right.\nonumber\\
&&\left.+\iota y\left\{\rho\gamma^2r
-\lambda^2s\right\}+\rho\gamma^2[u'(1+\gamma^2
u^2)(1+4\gamma^2 u^2)\right.\nonumber\\
&&\left.+2uV\gamma^2\{m'+2\gamma^2u^2V'\}]-s\lambda\lambda'
u\right]+sc_4\left\{\iota k_z\lambda
(1-u^2)\right.\nonumber\\
&&\left.+\lambda'-\lambda uu'\right\}+sc_5\{2\lambda uV\iota
k_z+\lambda um'-\iota k_x\lambda(1-V^2)]=0,
\end{eqnarray}
where
$m=V-\beta,~q=1+\gamma^2V^2,~r=1+\gamma^2u^2,~s=\frac{B^2}{4\pi},
~w=\omega+\beta k_x~\textrm{and}~y=k_xV+k_zu$. Equation (\ref{c4})
gives the relation between $k_x$ and $k_z$, i.e.,
$k_x=-\frac{c_5}{c_4}k_z$. We assume $c_4=c_5$ yielding $k_x=-k_z$
which reduces our wave vector to $\textbf{k}=(k_x,0,-k_x)$ with wave
number $k=|\textbf{k}|=\sqrt{2k_x^2}$. This assumption will make the
set of equations easier to solve. The relaxation in this assumption
may change the regions of normal and anomalous dispersion.

\section{Numerical Solutions}

When we solve the determinant of the coefficients of Eqs.(4.1),
(\ref{d4}), (\ref{e4})-(\ref{b4}) of constants
($c_1,~c_2,~c_3,~c_4,~c_5$) for the $x$-component of the wave
vector, it gives a complex dispersion relation of the form:
\begin{eqnarray}{\setcounter{equation}{1}}\label{dr}
\label{r1} &&A_1(\omega,z){k_x}^4+A_2(\omega,z){k_x}^3
+A_3(\omega,z){k_x}^2+A_4(\omega,z)k_x+A_{5}(\omega,z)\nonumber\\
&&+\iota\{B_1(\omega,z){k_x}^5+B_2(\omega,z){k_x}^4+B_3(\omega,z){k_x}^3
+B_4(\omega,z){k_x}^2\nonumber\\
&&+B_5(\omega,z)k_x+B_6(\omega,z)\}=0.
\end{eqnarray}
This relation is quintic in $k_x$ and cannot give exact solutions.
Thus, we solve it numerically by using \emph{Mathematica}, for the
complex values of $k_x$. The assumption $k_x=-k_z$ implies that
$k_z$ is also a complex number. Thus the quantities given by
Eq.(\ref{ps}) take the form
$$\sim e^{-\iota(\omega t-k_x x-k_z z)}=e^{-\iota(\omega t-k_{Rx}
x-k_{Rz} z)-k_{Ix} x-k_{Iz} z},$$ where $k_x=k_{Rx}+\iota k_{Ix}$
and $k_z=k_{Rz}+\iota k_{Iz}$. The values $k_{Rx}$ and $k_{Rz}$
represent the $x$ and $z$-components of the propagation factor
from which we can obtain the $x$ and $z$-components of the phase
and group velocities whereas $k_{Ix}$ and $k_{Iz}$ represent the
$x$ and $z$-components of the attenuation factor.

We have considered the region $-5\leq z\leq5$ to analyze the wave
properties of the cold plasma. The cold flow assumption indicates that the fluid has no viscosity
and no heat conduction. This is indeed an ideal case but provides basis for
the case of hot plasma, thus has its own importance. Perturbations
themselves modify this assumption because we include the perturbations
in the GRMHD equations for cold plasma. Since the flow variables
$u,~V,~\lambda$ have more variations near the pair production
region, thus we have excluded the region $-1<z<1$ and solved the
relation for rest of the region where these variables have small
variation. Thus the region under consideration $-5\leq z\leq5$ is
divided into two regions $-5\leq z\leq-1$ and $1\leq z\leq5$ for
which roots are numerically interpolated. The region $-5\leq
z\leq-1$ indicates the neighborhood of the pair production region
towards the event horizon and the region $1\leq z\leq5$ shows the
neighborhood of the pair production region towards the outer end
of the magnetosphere. We take the step-length $0.2$ for $z$ and
$\omega$ and find the value of $k_x$ at each point
$(z_i,\omega_j)~(i=1,...,21,~j=1,...51)$ of the mesh. Then by
separating each root, we estimate the surface by numerical
interpolation. The $x$-components of the propagation vector,
attenuation vector, phase and group velocities can be evaluated
from these interpolation functions.

We have obtained two velocities (given by Eqs.(\ref{u1}) and
(\ref{u2})) for our assumed plasma (i.e., rotating cold plasma with
constant rest-mass density). For each velocity, Eq.(\ref{dr}) leads
to five values of the $x$-component of the wave numbers. We display
$x$-components of the propagation vector, attenuation vector, phase
and group velocities and analyze the dispersion with respect to
these quantities.

For the region towards the black hole horizon with the fluid flow
velocity given by Eq.(\ref{u1}), we obtain one root for which the
wave number is infinite at several values of $z$. Such a root is
also found for the region towards the outer end of the magnetosphere
($z=\infty$) for the velocity components (Eq.(\ref{u2})). These
roots indicate that the waves are evanescent there which decreases
with the passage of time. Thus the energy cannot pass through the
region via waves and hence these cases are not interesting. For
the velocity components (\ref{u1}), the dispersion relations
obtained in the region towards the event horizon in the neighborhood
of the pair production region are given by Figures 1-4. Figures 6-9
indicate the dispersion relations for the region towards the outer
end of the magnetosphere. The dispersion relations related to the
velocity components (\ref{u2}) for the region towards the event
horizon are given by Figures 10-14. For the region towards the outer
end of the magnetosphere, these relations are shown by Figures
15-18.

In Figures 1-4 and 10-14, the positive propagation factor shows that
the waves are moving towards the pair production region whereas the
negative propagation factor indicates that the waves are moving
towards the event horizon. We see from Figures 6-9 and 15-18 that
when the propagation factor is positive, the waves are moving
towards the outer end of the magnetosphere whereas the negative
propagation factor indicates that the waves are moving towards the
pair production region. Table 1 indicates the regions of
magnetosphere with the directed waves. The rest of the region
contains random points of positive and negative propagation factor
and thus the direction of waves changes at each point.

If the attenuation factor is increasing, it leads to the damping
of waves whereas if this factor decreases, it indicates wave
growth. Table 2 shows the damping and growth of waves. The
up-arrow in the table shows the increase whereas the down-arrow
shows the decrease in respective quantity. The attenuation factor
takes random values otherwise.

In a region where the phase velocity is greater than the group
velocity, the waves are dispersed normally \cite{Ac}-\cite{Pain}
whereas the group velocity greater than the phase velocity shows
that the waves disperse anomalously. Table 3 shows the regions of
normal and anomalous dispersion of waves in respective figures.
Rest of the region contains random points of normal and anomalous
dispersion.

We note that in all the tables star (*) means validity of the
attribute to the whole region except for waves with negligible
angular frequency.

\begin{table}
\begin{tabular}{|c|c|c|c|c|}
  \hline
  Figure & Region of waves & Region of waves & Region of waves \\
  No. & moving towards & moving towards &  moving towards \\
    & the event horizon &  pair production region & the outer end \\\hline
  1 & --- & $-2.62\leq z\leq-1$* & --- \\\hline
  2 & $-2\leq z\leq-1$ & --- & --- \\\hline
  3 & --- & $-2\leq z\leq-1$ & --- \\\hline
  4 & --- & $-2.1\leq z\leq-1$& --- \\\hline
  5 & --- & Entire Region & --- \\\hline
  6 & --- & --- & Entire Region* \\\hline
  7 & --- & --- & Entire Region* \\\hline
  8 & --- & --- & Entire Region* \\\hline
  9 & --- & --- & Entire Region* \\\hline
  10 & Entire Region & --- & --- \\\hline
  11 & --- & Entire Region & --- \\\hline
  12 & --- & Entire Region* & --- \\\hline
  13 & --- & Entire Region* & --- \\\hline
  14 & --- & Entire Region & --- \\\hline
  15 & --- & $1\leq z\leq2.7$ & --- \\\hline
  16 & --- & --- & $1\leq z\leq 2.25$* \\\hline
  17 & --- & $1\leq z\leq2$* & --- \\\hline
  18 & --- & --- & Entire Region \\ \hline
\end{tabular}
\caption{Table indicating the direction of waves in respective
regions.}
\end{table}

\begin{table}
\begin{tabular}{|c|c|c|c|c|}
  \hline
  Figure & Regions of & Regions of \\
   No. & Wave Growth & Wave Damping \\\hline
  1 & $-1.6\leq z\leq-1,$*& \\
    & $0.398\leq\omega\leq7.25$ & --- \\
    &  as $\omega \uparrow$ and $z \downarrow$ & \\\hline
  2 \& 3 & Random & Random \\\hline
  4 & --- & $-1.5\leq z\leq-1$ as $\omega \uparrow$ and $z \downarrow$ \\\hline
    & $0\leq \omega\leq 0.23$ as $\omega \uparrow$ & $1.4\leq z\leq 5$ as $z \uparrow$ \\
  5 & $1\leq z\leq 1.4,~0.23\leq\omega\leq10$ & $1\leq z<1.4,~0\leq\omega\leq 0.23$ \\
    & as $z \uparrow$ &  as $z \uparrow$ \\ \hline
    &  & $1\leq z\leq2,~0.35\leq\omega\leq10$ \\
  6 & $\underline{~~~}$ &  as $z \uparrow$ and $\omega \downarrow$ \\
    &  & $2\leq z\leq5,~0.35\leq\omega\leq10$ \\
    &  &  as $z \uparrow$ \\\hline
  7 & --- & As $z \uparrow$* \\\hline
  8 & --- & As $\omega \uparrow$ and $z \downarrow$ \\\hline
    & $1.425< z\leq5,~0.2\leq\omega\leq10$ & \\
  9 & as $z \uparrow$ & $1\leq z\leq1.425,~0.2\leq\omega\leq10$ \\
    & $1\leq z\leq5,~0\leq\omega<0.2$ & as $z \uparrow$ \\
    & as $z \uparrow$ & \\ \hline
     & $-1.35\leq z\leq-1$ as $|z| \uparrow$ & \\
  10 & $1\leq z\leq2,~0.195\leq\omega\leq10$ & $-5\leq z\leq -1.35$ as $|z| \uparrow$ \\
     &  as $z \uparrow$ & \\ \hline
  11 & $-1.35\leq z\leq-1$ as $z \downarrow$ & $-5\leq z\leq-1.35$ as $z \downarrow$ \\\hline
  12 & As $|z|\uparrow$ &  As $\omega \uparrow$ \\ \hline
  13 & $z \downarrow$ and $\omega \uparrow$ & --- \\ \hline
  14 & $\underline{~~~}$ & $-5\leq z\leq-1,0.175\leq\omega\leq10$ \\
     &  &  as $z \uparrow$ and $\omega \uparrow$ \\\hline
     &  & $1\leq z\leq2,~0.195\leq\omega\leq10$ \\
     &  &  as $z \uparrow$ \\
  15 &  $1.6\leq z\leq1.8,~0\leq\omega\leq0.195$ & $1\leq z\leq1.6,~0\leq\omega\leq0.195$ \\
     &  as $z \uparrow$ and $\omega \uparrow$ &  as $z \uparrow$ and $\omega \uparrow$ \\
     &  & $1.8\leq z\leq2,~0\leq\omega\leq0.195$  \\
     &  & as $z \uparrow$ and $\omega \uparrow$ \\\hline
  16 & $1.5\leq z\leq1.625$ & $1.0\leq z\leq1.5$ \\\hline
  17 \& 18 & Random & Random \\\hline
\end{tabular}
\caption{Table indicating the regions of growth and damping of
waves.}
\end{table}

\begin{table}
\begin{tabular}{|c|c|c|c|c|}
   \hline
  Figure & Regions of  & Regions of  \\
  No. & Normal Dispersion & Anomalous Dispersion \\ \hline
  1 & $-1.6\leq z\leq-1,~0.1\leq\omega\leq7.25$ & --- \\ \hline
  2 & --- & $-2.15\leq z\leq-1,~0<\omega\leq 0.05$ \\ \hline
  3 & Random & Random \\ \hline
  4 & --- & $-2\leq z\leq-1$ \\ \hline
  5 & $1\leq z\leq1.25,~0<\omega\leq 2$ & \\
    & $1.25\leq z<3,~0<\omega\leq 1$ & --- \\
    & $3\leq z\leq5,~0\leq\omega\leq0.75$ & \\ \hline
  6 &  & $2\leq z\leq5,~0.8\leq\omega\leq1$, \\
    & --- & $1\leq z\leq4,~0.5\leq\omega\leq0.8$, \\
    &  & $1\leq z\leq5,~0<\omega\leq0.5$  \\ \hline
  7 & Entire region* & --- \\ \hline
  8 & --- & $0.2225\leq \omega\leq5$ \\\hline
  9 & $\omega<0.4025$ & --- \\\hline
 10 & $-5\leq z\leq-4.2,~0<\omega\leq0.1$ & --- \\\hline
 11 & $-3\leq z\leq-2.5,~0<\omega\leq0.4$ & $-5\leq z\leq-3,~0<\omega\leq5$\\
    &  & $-2.5\leq z\leq-1,~0<\omega\leq10$\\\hline
 12 & Random & Random \\\hline
 13 & $-5\leq z\leq-2,~0<\omega\leq0.2$ & $-5\leq z\leq-1,~0.2\leq\omega\leq10$ \\
    &  & $-2\leq z\leq-1,~0.16\leq\omega\leq0.2$ \\\hline
    & $-2\leq z\leq-1,~0.1\leq\omega\leq0.39$ & \\
 14 & $-2\leq z\leq-1,~0.41\leq\omega\leq0.5$ & --- \\
    & $-2.2\leq z\leq-1,~0<\omega\leq0.08$ & \\\hline
 15 & --- & $1.8\leq z\leq2,~0.002\leq\omega\leq0.003$ \\\hline
    & $1\leq z\leq1.4,~2.5\leq\omega\leq10$ & \\
 16 & $1\leq z\leq1.4,~1\leq \omega\leq1.6$ & --- \\
    & $1\leq z\leq1.25,~1.6\leq\omega\leq2$ & \\\hline
 17 & $1.8\leq z\leq2,~0\leq\omega\leq0.00015$ & $1\leq z\leq1.5,~0\leq\omega\leq1.25$ \\\hline
 18 & $1\leq z\leq1.45,~0<\omega\leq3.5$ & --- \\\hline
\end{tabular}
\caption{Table indicating the regions admitting normal and
anomalous dispersion of waves.}
\end{table}

\section{Conclusion}

This work is devoted to investigate the wave properties in the
neighborhood of the pair production region. We have considered
rotating cold plasma filled magnetosphere. The perturbations are
assumed to be simple harmonic waves caused by the gravity
influence as well as the rotation of the black hole. The
determinant of the Fourier analyzed GRMHD equations is solved to
obtain the $x$-component of the wave vector which is found to be a
complex number. This has been done numerically for the two values
of the velocity of the magneto-rotational fluid obtained in
Section \textbf{3}. The relation between the $x$ and
$z$-components of the wave vector is $k_x=-k_z$. This indicates
that if we can have information about $k_x$ and its corresponding
quantities, we can infer the results for the waves moving along
$z$-axis.

The summary of these results can be expressed as follows:

The propagation and attenuation factors take random values far
from the pair production region in Figures 1-4, 15-18. The
propagation factor increases with increasing angular frequency in
Figures 6-9, 11-14. This gives the increment in propagation of
waves along $x$-axis with the increment in angular frequency. It
is observed in Figures 5-14 that the phase and group velocities
take their extreme values (either maximum or minimum) near the
pair production region. Thus the pair production region allows the
waves to take extreme values near it.

In most of the figures, the region in the extreme neighborhood of
the pair production region admits the waves to pass through (normal
dispersion) which indicates that the waves can pass through this
region. Anomalous dispersion in the extreme neighborhood of the pair
production region in Figures 2, 4, 6, 8, 11, 13 and 17. In the region
of anomalous dispersion, the phase velocity is less than the
group velocity, so the waves cannot carry energy alongwith. Most of
the figures show random dispersion of waves in the far regions of the
pair production region.

Figure 7 shows normal dispersion of waves throughout the region
which indicates that the waves can move towards the outer end of
the magnetosphere. This verifies the result given by Zhang
\cite{Z2} and Sheikh \cite{S10} that cold plasma allows outflow of
energy flux from the pair production region.

It would be worth exploring to check these properties by taking a
hot plasma in a rotating magnetosphere.

\begin{figure}
\center\epsfig{file=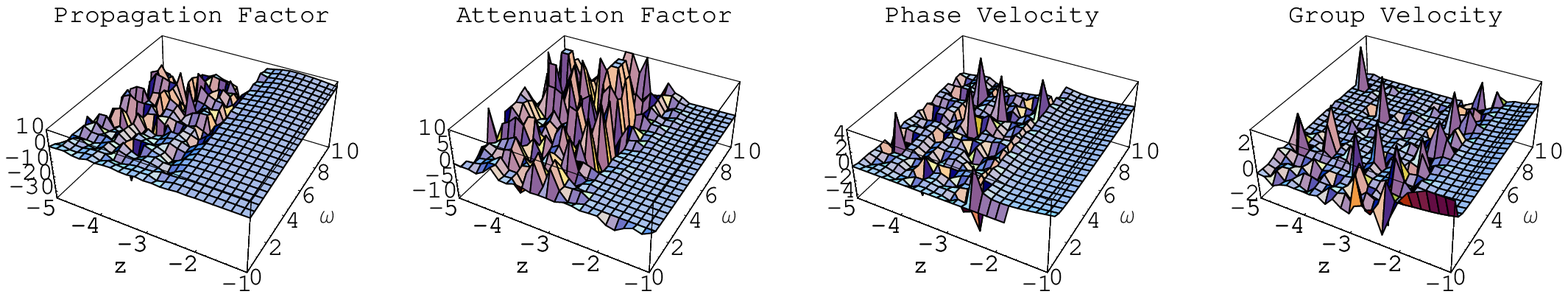,width=1.0\linewidth} \caption{A large
region admits normal dispersion of waves moving towards the pair
production region.}
\center\epsfig{file=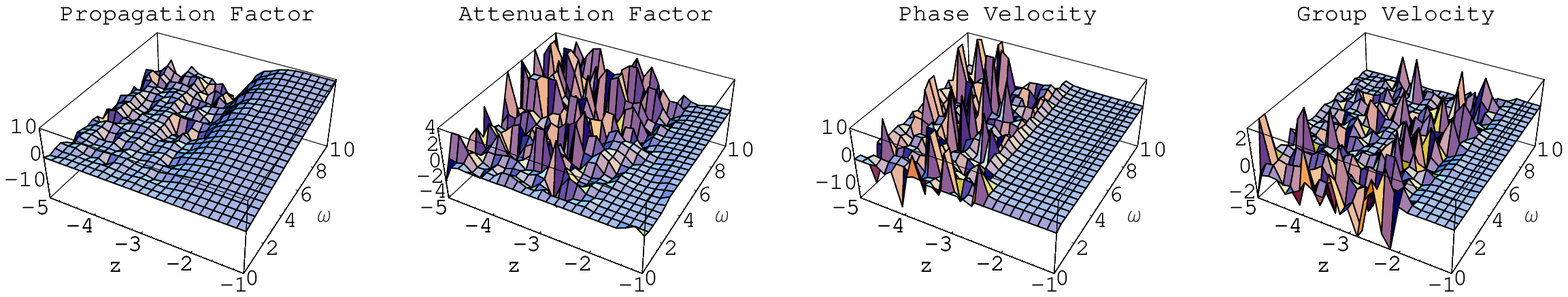,width=1.0\linewidth} \caption{A small
region in the neighborhood of the pair production region indicates
anomalous dispersion at small angular frequencies. A small region
shows the waves are moving towards the event horizon.}
\center\epsfig{file=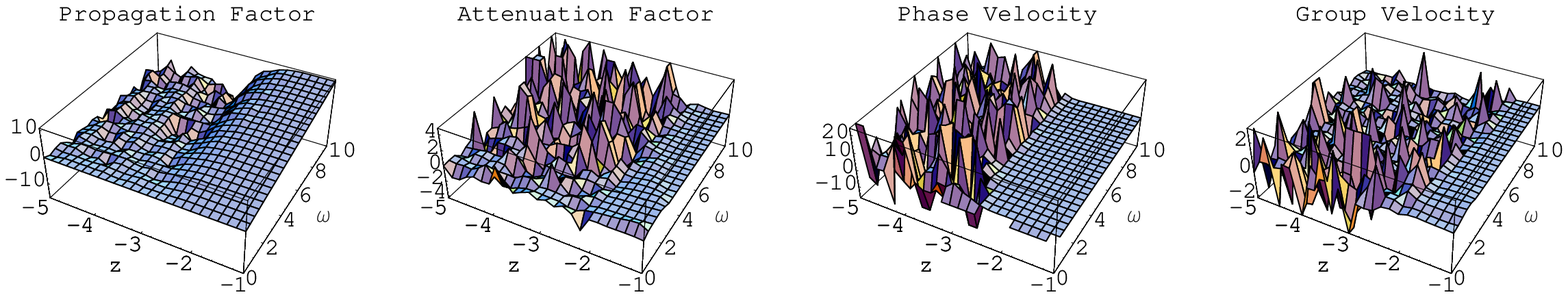,width=1.0\linewidth} \caption{Random
dispersion of waves is found. A small region shows that the waves
are moving towards the pair production region.}
\center\epsfig{file=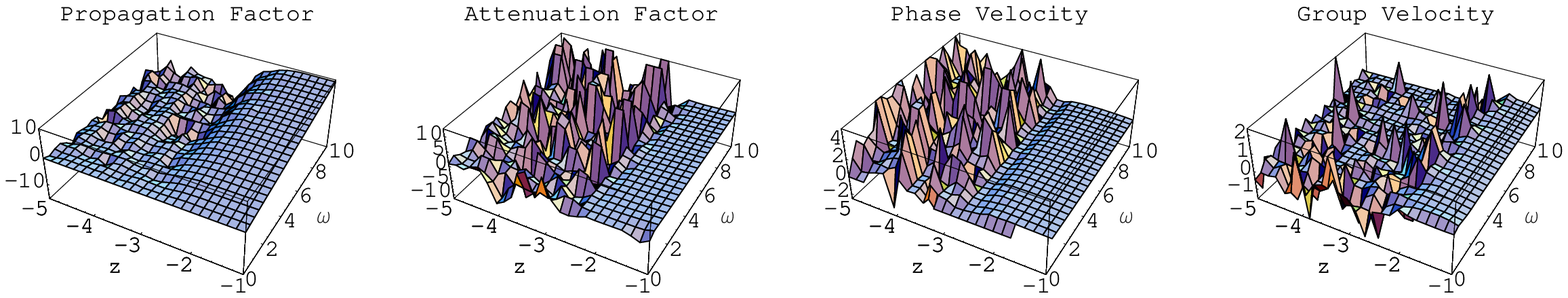,width=1.0\linewidth} \caption{The
region near to the pair production region shows anomalous
dispersion for the waves moving towards the pair production
region.}
\end{figure}
\begin{figure}
\center \epsfig{file=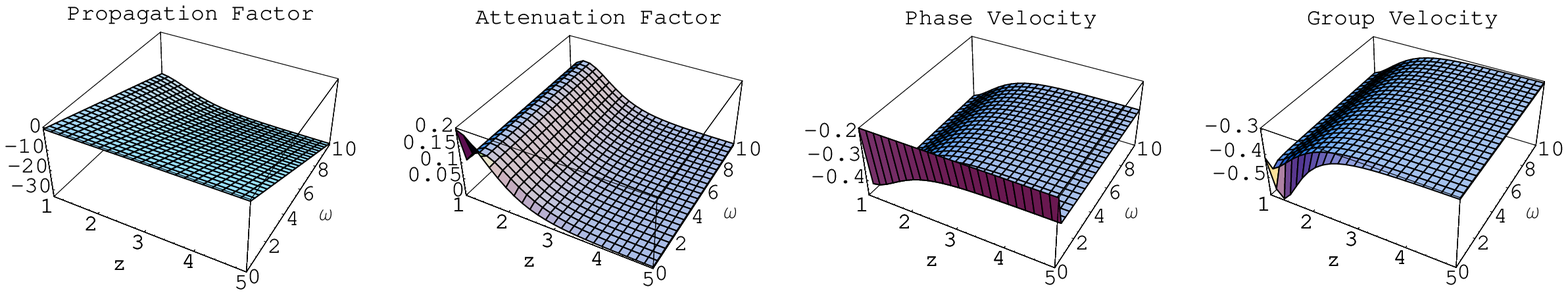,width=1.0\linewidth} \caption{Wave
propagation decreases as the waves move towards the pair
production region. Small regions admitting normal dispersion are
found.} \center \epsfig{file=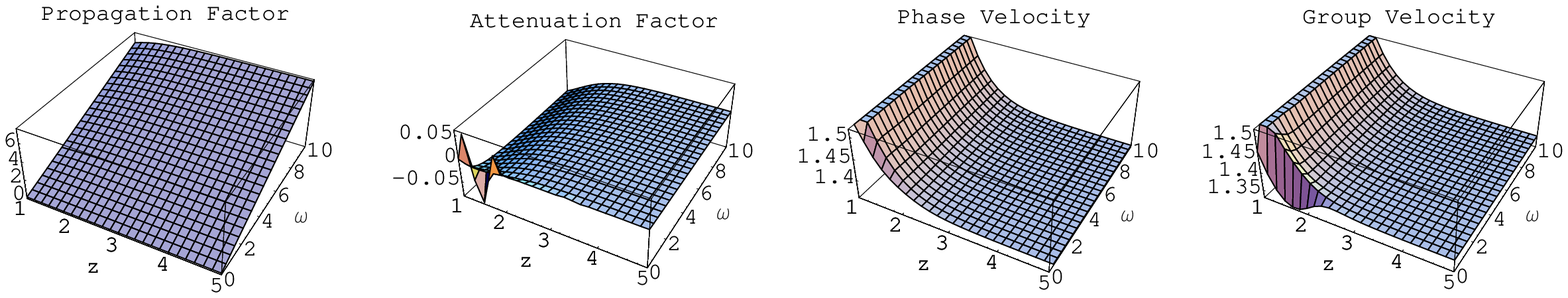,width=1.0\linewidth}
\caption{The waves damp as they move away from the pair production
region except for the waves with very small angular frequencies.
Regions of anomalous dispersion are found for small angular
frequencies in the neighborhood of the pair production region.}
\center \epsfig{file=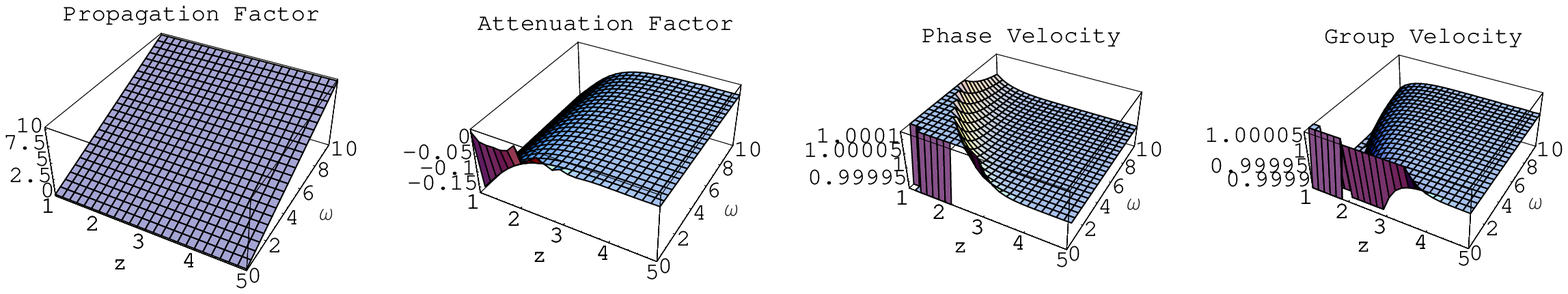,width=1.0\linewidth} \caption{The
waves damp while moving away from the pair production region
(except for the waves with small angular frequencies). Normal
dispersion is found except for the negligible angular frequency
waves.}\end{figure}
\begin{figure}
\center\epsfig{file=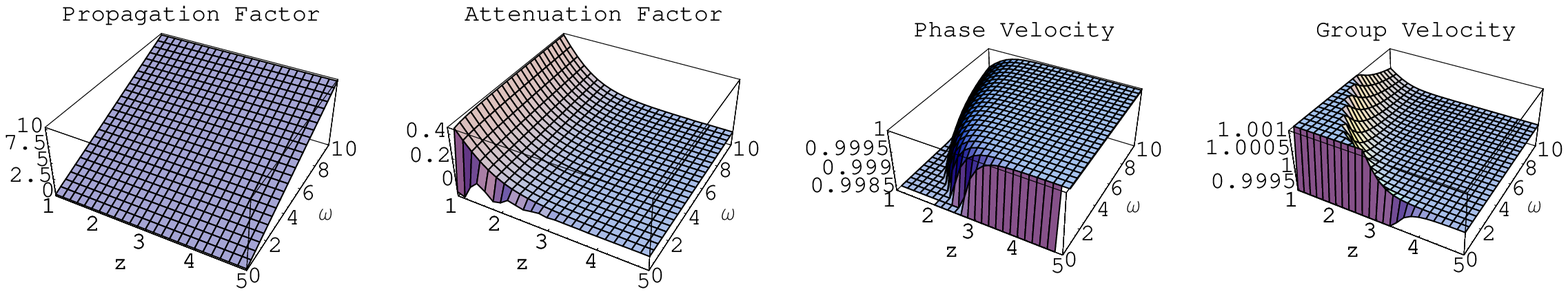,width=1.0\linewidth} \caption{Waves
grow as they move away from the pair production region. In most of
the region, dispersion is anomalous except for the waves with very
small angular frequencies lying between $0.2225$ and $5$.}\center
\epsfig{file=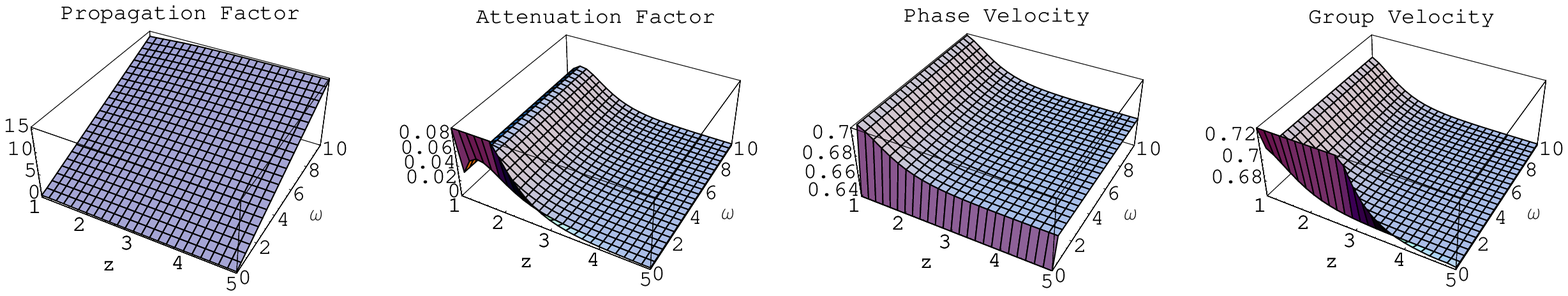,width=1.0\linewidth} \caption{The propagation
of waves increases with increasing angular frequency. Waves with
small angular frequencies admit normal dispersion.}
\center\epsfig{file=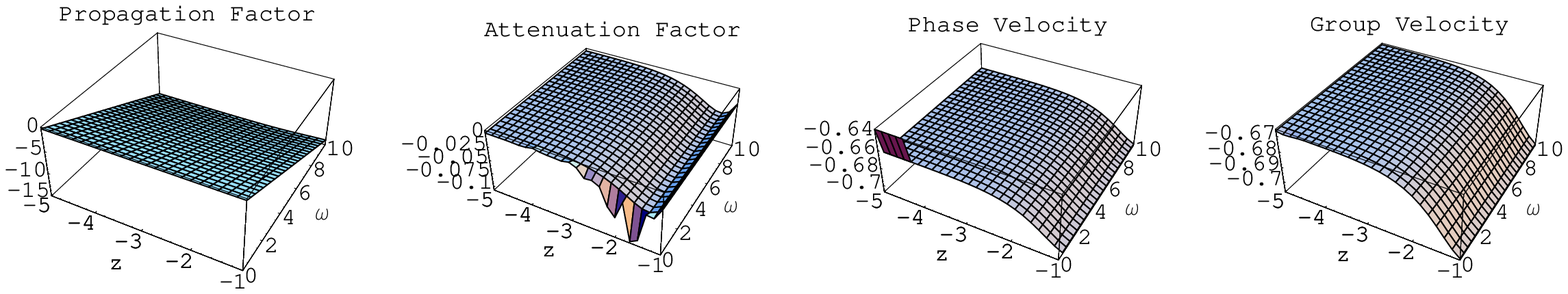,width=1.0\linewidth} \caption{Waves
grow and then damp when the value of $z$ increases. This
propagation decreases as their angular frequency increases.
Dispersion is found to be normal in a small region.}\end{figure}
\begin{figure}
\center\epsfig{file=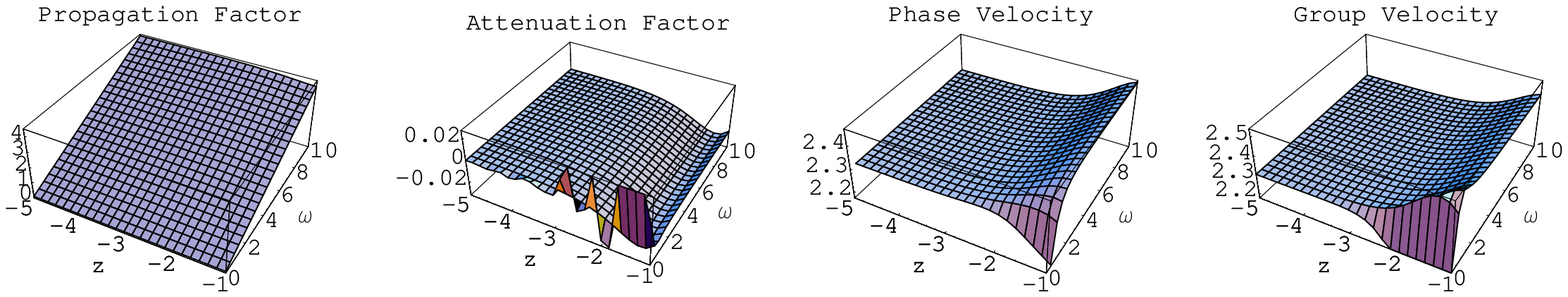,width=1.0\linewidth} \caption{The
propagation of waves decreases when the waves move away from the
event horizon. Waves damp first and then grow while moving towards
the pair production region. Small regions show normal and
anomalous dispersion of waves.}
\center\epsfig{file=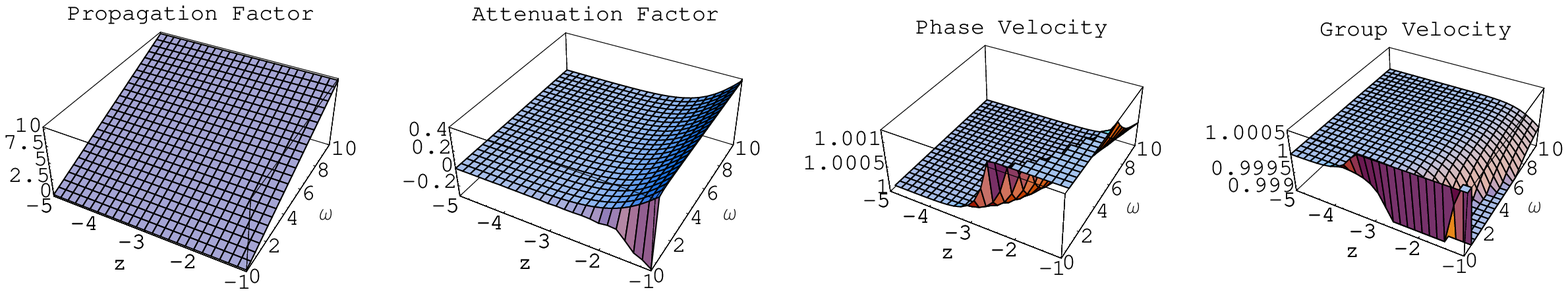,width=1.0\linewidth} \caption{Waves
damp as they move towards the pair production region. Random
points of normal and anomalous dispersion are found.}
\center\epsfig{file=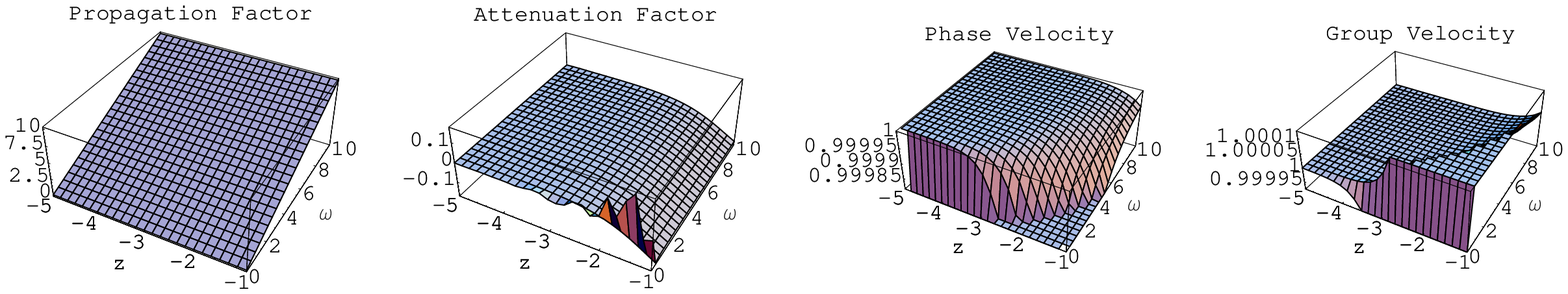,width=1.0\linewidth} \caption{The
increase in the angular frequency increases the wave propagation.
Waves grow as they move towards the pair production region. A
region of normal dispersion is found for small frequency waves.}
\center\epsfig{file=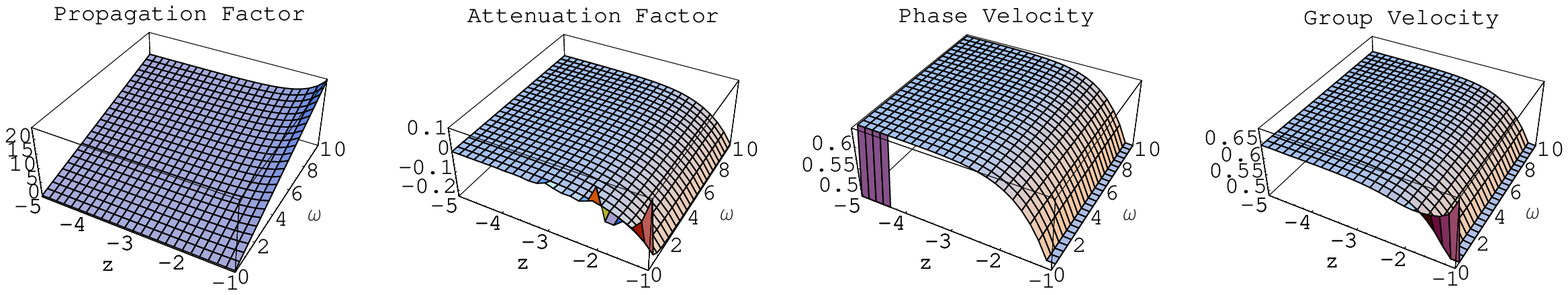,width=1.0\linewidth} \caption{The wave
propagation decreases and the waves damp as they move towards the
pair production region except for the low angular frequency waves.
Small region admitting normal dispersion of waves are identified.}
\end{figure}
\begin{figure}
\center\epsfig{file=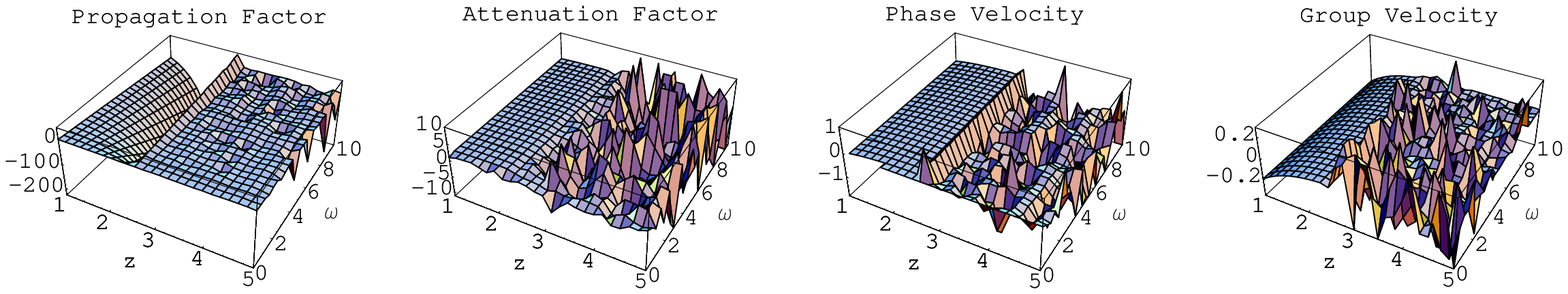,width=1.0\linewidth}\caption{The
medium admits a small region for the waves moving towards the pair
production region. The wave propagation decreases and then
increases as the waves move away from the event horizon. Small
regions of normal and anomalous dispersion are found.}
\center\epsfig{file=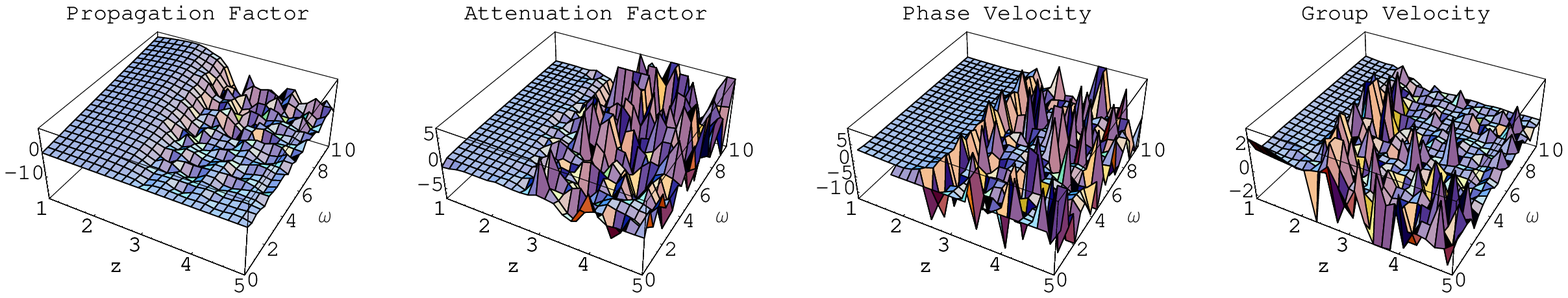,width=1.0\linewidth}\caption{The
propagation vector, attenuation factor, phase and group velocity
vectors admit random values away from the pair production region.
Small regions near the pair production region show normal
dispersion.}
\center\epsfig{file=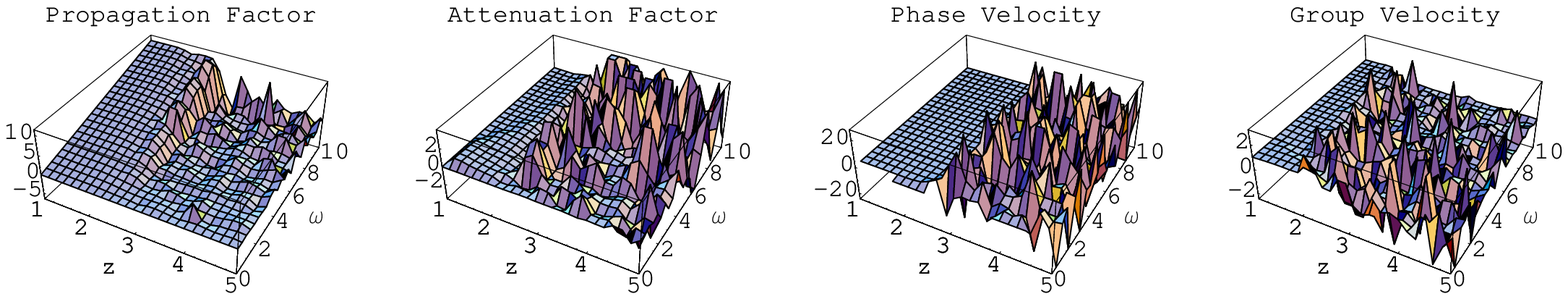,width=1.0\linewidth}\caption{A small
region admitting anomalous dispersion of waves moving towards the
outer end of the magnetosphere.}
\center\epsfig{file=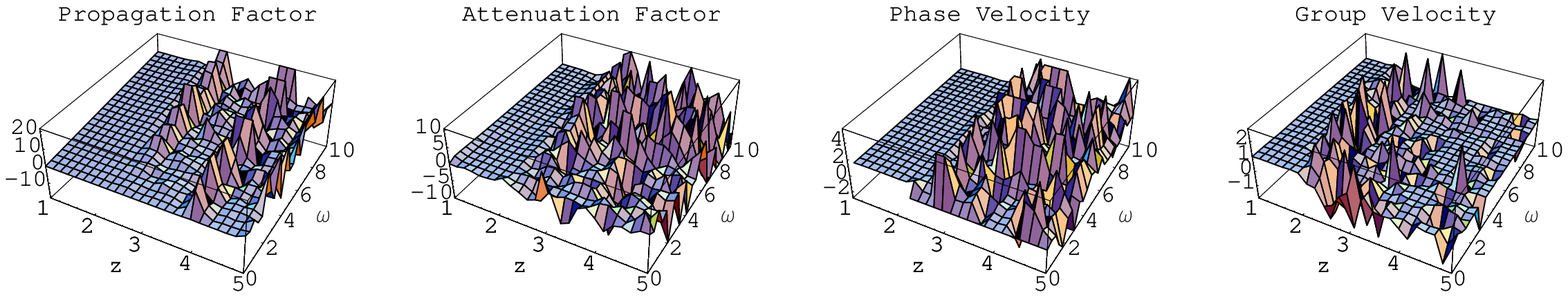,width=1.0\linewidth}\caption{A small
region in the neighborhood of the pair production region shows
normal dispersion of the low angular frequency waves moving
towards the outer end of the magnetosphere.}
\end{figure}

\vspace{0.5cm}

\begin{thebibliography}{99}
\bibitem{RW} Regge, T. and Wheeler, J.A.: Phys. Rev.
\textbf{108}(1957)1063.

\bibitem{Ze} Zerilli, F.: Phys. Rev. \textbf{D2}(1970)2141;
Phys. Rev. Lett. \textbf{24}(1970)737.

\bibitem{Pr} Price, R.H.: Phys. Rev. \textbf{D5}(1972)2419;
ibid 2439.

\bibitem{HR}Hanni R.S. and Ruffini, R.: Phys. Rev.
\textbf{D8}(1973)3259.

\bibitem{Wa}Wald, R.M.: Phys. Rev. \textbf{D10}(1974)1680.

\bibitem{M}Mashhoon, B.: Phys. Rev. \textbf{D10}(1974)1059.

\bibitem{SK}Sakai, J. and Kawata, T.: J. Phys. Soc. Jpn.
\textbf{49}(1980)747.

\bibitem{Go} Gosh, P: Mon. Not. R. Astron. Soc. \textbf{315}(2000)89.

\bibitem{MK}Moortgat, J. and Kuijpers, J.: Mon. Not. R. Astron. Soc.
\textbf{368}(2006)1110.

\bibitem{L} Lynden-Bell, D.:  Nature \textbf{223}(1969)690.

\bibitem{BZ} Blandford, R.D. and Znajek, R.L.: Mon. Not. R. Astron. Soc.
\textbf{179}(1977)433.

\bibitem{T1} Takahashi, M., Nitta, S., Tatematsu, Y. and Tomimatsu, A.:
Astrophys. J. \textbf{363}(1990)206.

\bibitem{ADM} Arnowitt, R., Deser, S. and Misner, C.W.: \textit{Gravitation:
An Introduction to Current Research} ed. Witten, L. (Wiley, New
York, 1962); gr-qc/0405109v1.

\bibitem{TM1} Thorne, K.S. and Macdonald, D.A.: Mon. Not. R. Astron. Soc.
\textbf{198}(1982)339; ibid \textbf{198}(1982)345.

\bibitem{TPM} \textit{Black Holes: The Membrane Paradigm}
eds. Thorne, K.S., Price, R.H. and Macdonald, D.A. (Yale
University Press, New Haven, 1986).

\bibitem{HT} Holcomb, K.A. and Tajima, T.: Phys. Rev. \textbf{D40}(1989)3809.

\bibitem{De} Dettmann, C.P., Frankel, N.E. and Kowalenko, V.:  Phys. Rev.
\textbf{D48}(1993)5655.

\bibitem{BHT1} Buzzi, V., Hines, K.C. and Treumann, R.A.: Phys. Rev.
\textbf{D51}(1995)6663; ibid 6677.

\bibitem{S1} Sharif, M. and Sheikh, U.: Gen. Relat. Gravit. \textbf{39}(2007)1437;
ibid 2095.

\bibitem{S3} Sharif, M. and Sheikh, U.: Int. J. Mod. Phys.
\textbf{A23}(2008)1417.

\bibitem{S4} Sharif, M. and Sheikh, U.: J. Korean Phys.
Soc. \textbf{52}(2008)152.

\bibitem{S5} Sharif, M. and Sheikh, U.: J. Korean Phys. Soc. \textbf{53}(2008)2198.

\bibitem{Ko1} Komissarov, S.S.: Mon. Not. R. Astron. Soc.
\textbf{336}(2002)759.

\bibitem{Kh} Khanna, R.: Mon. Not. R. Astron. Soc. \textbf{294}(1998)673.

\bibitem{Pen}  Penrose, R.: Riv. Nuovo Cimento \textbf{1}(1969)252.

\bibitem{LK} Leiter, D. and Kafatos, M.: Astrophys. J. \textbf{226}(1978)32.

\bibitem{Kaf} Kafatos, M.: Astrophys. J. \textbf{236}(1980)99.

\bibitem{Ko} Koide, S., Shibata, K., Kudoh, T. and Meier, D.L.:
Science \textbf{295}(2002)1688.

\bibitem{KS} Kokkotas, K. and Schmidt, B.: Liv. Rev. Rel \textbf{2}(1999)2.

\bibitem{FN} Furuhashi, H. and Nambu, Y.: Prog. Theor. Phys. \textbf{112}(2004)983-995.

\bibitem{Zeld} Zel'dovich, Y. B.: Zh. Eksp. Teor. Fiz. \textbf{62}(1972)2076 (Sov. Phys. JETP \textbf{35} (1972)1085).

\bibitem{Z1} Zhang, X.-H.: Phys. Rev. \textbf{D39}(1989)2933.

\bibitem{Z2} Zhang, X.-H.: Phys. Rev. \textbf{D40}(1989)3858.

\bibitem{S6} Sharif, M. and Sheikh, U.: J. Korean Phys. Soc. (2009, to appear).

\bibitem{S7} Sharif, M. and Sheikh, U.: Class. Quantum Grav.
\textbf{24}(2007)5495.

\bibitem{S9}  Sharif, M. and Sheikh, U.: Canadian J. Phys. (2009, to appear).

\bibitem{S10} Sheikh, U.: \textit{Ph.D. Thesis} (University of the Punjab, Lahore, 2007)

\bibitem{Ac} Achenbach, J.D.: \textit{Wave Propogation in Elastic Solids}
(North-Holland Publishing Company, Oxford, 1973).

\bibitem{Pain} Pain, H.J.: \textit{The Physics of Vibrations and
Waves} (John Wiley and Sons, Chichester, 2005).

\end{thebibliography}
\end{document}